\documentstyle[preprint,aps,floats,epsfig]{revtex}

\begin{document}

\draft

\preprint{\rightline{ANL-HEP-PR-98-51}}

\title{Improved staggered quark actions with reduced flavour symmetry 
       violations for lattice QCD}
\author{J.-F.~Laga\"{e} and D.~K.~Sinclair}
\address{HEP Division, Argonne National Laboratory, 9700 South Cass Avenue,
         Argonne, IL 60439, USA}

\maketitle

\begin{abstract}
We introduce a new class of actions for staggered quarks in lattice QCD which
significantly reduce flavour symmetry violations in the pion mass spectrum.
An action introduced by the MILC collaboration for the same purpose is seen to
be a special case. We discus how such actions arise from a systematic attempt
to reduce flavour symmetry violations in the weak coupling limit. It is shown
that for quenched lattice QCD at $6/g^2=5.7$, representative actions of this
class give a considerable reduction in flavour symmetry violation over the
standard staggered action, and a significant reduction over what is achieved by
the MILC action. 
\end{abstract}


\setcounter{page}{1}
\pagestyle{plain}
\parskip 5pt
\parindent 0.5in

\section{introduction}

Lattice QCD simulations have always been limited by the requirement that the
lattice be large compared with the correlation lengths (in particular with the
pion compton wavelength) of the theory, while the lattice spacing be small
enough for physical observables to exhibit the scaling properties dictated by
asymptotic freedom. In addition the symmetries of the lattice theory should
approximate the continuum Lorentz and flavour symmetries. With the standard
action this requires a lattice with a large number of sites. For this reason
there has been considerable effort to find improved actions which obtain the
desired results with appreciably larger lattice spacings.

Two methods have emerged for producing such improved actions. (For a recent
summary of such methods and a more complete set of references see
\cite{niedermayer}.) In the first method,
the action is improved in powers of the lattice spacing $a$, by adding higher
dimensional operators \cite{symanzik}. 
The coefficients can be calculated perturbatively \cite{lm} or
non-perturbatively \cite{luscher}. 
The second method uses renormalization group analyses to
determine the fixed point, ``perfect'' action \cite{hasenfratz} (This paper
references to the earlier literature).
For the pure gauge sector of
lattice QCD, actions of the L\"{u}scher-Weisz form \cite{lw} 
and truncated perfect actions \cite{hasenfratz} work very well. 
For the fermion part of the action, Sheikholeslami and
Wohlert have identified the operators which contribute at low orders in $a$
\cite{sw}.
For Wilson fermions both methods based on expansions in $a$ and on perfect
actions have had considerable success with lattice spacings as large as 
$\sim 0.5$~fm \cite{alford,degrand}.

For staggered fermions, the improvement of the action in powers of $a$ was
discussed by Naik \cite{naik}. 
More recently Luo has enumerated all operators which 
contribute at low orders in $a$ \cite{luo}. 
Perfect action methods have also been 
applied to staggered fermions \cite{bbcw,bd}. 
While simple actions of both these classes show
improvements over the standard action, they do not significantly reduce the
flavour symmetry violations in the pion spectrum \cite{milcs,bielefeld}. 
These flavour symmetry 
violations are a major part of the reason one is forced to use small lattice
spacings when using staggered fermions.

The reason why these simple improvements to the staggered quark action fail to
make significant improvements to the flavour symmetry, is that they concentrate
on making the fermion dispersion relations near the corners of the Brillouin
zone (where the components of the fermion momentum are close to $0$ or $\pi$ in
lattice units) more continuum-like, and the interactions of these fermions with
low momentum gauge fields also more continuum-like. However, the reason for
the flavour symmetry violations in the staggered fermion method is that the
gauge fields can transfer momenta large enough to take the quark from the
neighbourhood of one corner of the Brillouin zone to the vicinity of another,
which changes that quark's flavour. Such large momentum transfers are not
suppressed by these simple improved actions, and so flavour symmetry violations
are not significantly suppressed.

Recently, the MILC collaboration introduced an new staggered fermion action in
which the gauge link is replaced by the linear combination of a single gauge
link and the sum over the six 3-link ``staples'' joining the same sites
\cite{milc}. Such
a replacement tends to smooth the interaction between the gauge and fermion
fields, thus reducing the coupling of the high-momentum components of the gauge
fields to the quarks. Not surprisingly this action significantly  decreases the
flavour symmetry violation in the pion spectrum.

We have examined how to systematically suppress such flavour symmetry 
violations. At tree level, this leads to an action which reduces the flavour
symmetry breaking by an extra power of $a^2$. We then assume that a good choice
of action beyond tree level will have the same form, but with different
coefficients. The MILC action is then seen to be a special case of this more
general class of actions.

We have compared the spectrum of light hadrons, with particular emphasis on
the pions, obtained with the standard staggered quark action, the MILC action,
and a subset of our new actions. For a preliminary search of the parameter
space for these actions we calculated the spectrum of light hadrons on a set 
of quenched gauge configurations at $\beta=6/g^2=5.7$ on an $8^3 \times 32$
lattice. A summary of these results was presented at Lattice'97 \cite{ls}. 
We have since 
calculated the spectrum of light hadrons for the standard action, a near
optimal MILC action and a promising choice from our new actions on quenched
gauge configurations with $\beta=6/g^2=5.7$ on a $16^3 \times 32$ lattice. From
these calculations we conclude that, for an appropriate choice of parameters,
our new class of actions represents a significant improvement over the MILC
action, and confirm that both actions represent a considerable improvement over
the standard staggered action. 

In section 2 we describe how to systematically reduce flavour symmetry breaking
at tree level for lattice QCD at weak coupling, and introduce our new class of
improved actions based on this analysis. In section 3 we present our
measurements of the hadron spectrum on quenched configurations. Section 4 gives
our summary and conclusions.

\section{Improving the staggered quark action}

In the staggered fermion transcription of quarks to the lattice, the quark
field on site n of the lattice, $\psi(n)$, is a 3 component object --- a colour
triplet. It lacks Dirac or flavour indices. The 4 flavours and 4 Dirac 
components are associated with the 16 poles (per colour) of the free lattice 
Dirac propagator. For massless quarks, these occur when each component of
the momentum $p_\mu = 0$ or $\pi$. Since interactions, in general, change the 
momenta of the quarks, they induce mixings between the degrees of freedom
associated with different flavours and hence break flavour symmetry.

At tree level, one can suppress flavour mixing to higher order in $a$, if one
suppresses the coupling of fermions to gluons whose momentum components are
all either $0$ or $\pi$ but not all $0$. To see how this might be done, let us
for the moment ignore the requirements of gauge invariance. Then the quark
gluon coupling term in the Lagrangian could be replaced by
\begin{equation}
i \psi^{\dagger}(n) \eta_\mu(n) A_\mu(n) \psi(n+\mu) - h.c..
\end{equation}
which again gives mixing between quark flavours. We now replace $A_\mu$ in
this term by
\begin{equation}
A_\mu(n) \rightarrow \frac{1}{256}(2+D_1+D_{-1})(2+D_2+D_{-2})
                                  (2+D_3+D_{-3})(2+D_4+D_{-4}) A_\mu(n)
\label{eqn:naive}
\end{equation} 
where
\begin{equation}
D_{\pm\nu} A_\mu(n) = A_\mu(n\pm\nu).
\end{equation}
In momentum space this is equivalent to the substitution
\begin{equation}
A_\mu(k)  \rightarrow \frac{1}{16}(1+\cos k_1)(1+\cos k_2)
                                  (1+\cos k_3)(1+\cos k_4) A_\mu(k).
\end{equation}
The right hand side of this equation $\rightarrow A_\mu(k)$ as $k \rightarrow 0$
and differs from $A_\mu(k)$ by a factor of only 
$1+{\cal O}(a^2)$ for $|k| = {\cal O}(a)$. It vanishes when any component of
$k$ equals $\pi$ and is suppressed by a factor of ${\cal O}(a^2)$ when any
component of $k$ is within ${\cal O}(a)$ of $\pi$. Hence this modification of
the action would suppress the tree level flavour symmetry violations by
${\cal O}(a^2)$, which is what we want for our improved action.

We now return to the gauge invariant theory. The quark-gluon coupling term in
the Lagrangian is now
\begin{equation}
\psi^{\dagger}(n) \eta_\mu(n) U_\mu(n) \psi(n+\mu) - h.c..
\end{equation}
where $U_\mu = \exp i A_\mu$.
Our ansatz for a tree-level improved action is obtained by replacing $A_\mu$
by $U_\mu$ in equation~\ref{eqn:naive}, multiplying out the prefactors, and
replacing the products of displacement operators by appropriately symmetrized
covariant displacement operators. This leads to the replacement
$U_\mu \rightarrow {\bf U}_\mu$ where 
\begin{eqnarray}
{\bf U}_\mu = \frac{1}{16} \{ 2 
  + \sum_\nu [\frac{1}{2} D_\nu &+& \frac{1}{4} (D_{\mu\nu}+D_{-\mu\nu})] 
  + \sum_{\nu\rho} [\frac{1}{4} D_{\nu\rho} + \frac{1}{8} (D_{\mu\nu\rho}
                                     + D_{-\mu\nu\rho})]  \nonumber \\
&+& \sum_{\nu\rho\lambda} [\frac{1}{8} D_{\nu\rho\lambda}
  + \frac{1}{16} (D_{\mu\nu\rho\lambda}+D_{-\mu\nu\rho\lambda})]\} U_\mu
\label{eqn:tree}
\end{eqnarray}
where $\nu$, $\rho$ and $\lambda$ are summed over $\pm 1$, $\pm 2$, $\pm 3$,
$\pm 4$ with $|\mu|$, $|\nu|$, $|\rho|$, $|\lambda|$ all different. The $D$'s
are the covariant displacement operators. For example $D_\nu$ is defined by 
\begin{equation}
D_{\nu} U_\mu(n) = U_\nu(n) U_\mu(n+\nu) U^{\dagger}_\nu(n+\mu).
\end{equation}
The operational definition of the $D$'s is as follows. The link is displaced
one unit in each of the subscript directions. The ends of the undisplaced
link are joined to the ends of the displaced link by products of links over 
the shortest paths joining the two. We then symmetrize over all such paths.

We now must check that this replacement suppresses the quark-gluon interaction
when at least one component of the gluon momentum is close to $\pi$ and the
rest are near to $0$. Since we are interested primarily in what happens to
leading order in $a$ we can write $U_\mu \approx 1 + iA_\mu$ and
${\bf U}_\mu \approx 1 + i{\bf A}_\mu$. We then evaluate ${\bf A}_\mu$ when
each component of the momentum $k$ of $A_\mu$ is $0$ or $\pi$. When
$k_1=k_2=k_3=k_4=0$ we find ${\bf A}_\mu = A_\mu$. At the other corners of
the Brillouin zone we find that ${\bf A}_\mu$ is longitudinal, and hence
decouples, which is the desired result.
 
To go beyond tree level we assume that we can use a replacement of the same
form as the tree replacement, i.e.
\begin{eqnarray}
{\bf U}_\mu = C \{ x_0+2 y_0 
  + \sum_\nu [x_1 D_\nu &+& y_1 (D_{\mu\nu}+D_{-\mu\nu})] 
  + \sum_{\nu\rho} [x_2 D_{\nu\rho} + y_2 (D_{\mu\nu\rho}
                                     + D_{-\mu\nu\rho})]  \nonumber \\
&+& \sum_{\nu\rho\lambda} [x_3 D_{\nu\rho\lambda}
  + y_3 (D_{\mu\nu\rho\lambda}+D_{-\mu\nu\rho\lambda})]\} U_\mu
\label{eqn:improved}
\end{eqnarray}
with 
$ C = 1/( x_0 + 2 y_0 + 6 x_1 + 12 y_1 +  12 x_2 + 24 y_2 + 8 x_3 + 16 y_3 ) $
Here we could modify the definitions of the displacement operators $D$ which
include the index $\pm\mu$ to use different weights, depending on the positions
of the links in the $\pm\mu$ directions. Indeed standard tadpole improvement
\cite{lm}
would require such changes. We have chosen not to exercise this option in our
choices of improved actions, since even without, this class of action has 6
free parameters. We note that the MILC action belongs to this class, being the
special case where $x_0=1$, $x_1=\omega$, and $x_2=x_3=y_0=y_1=y_2=y_3=0$.

\section{The hadron spectrum with improved gauge actions.}

One of the most visible effects of flavour symmetry violation for staggered
quarks is seen in the pion mass spectrum. Only one of the pions is a true
Goldstone boson whose mass vanishes as the quark mass is taken to zero. The
mass differences between the non-Goldstone pions is typically somewhat less
than that between them and the Goldstone pion. For our measurements, we have
chosen $\pi_2$, the other local pion, as our representative non-Goldstone pion.
For inverse lattice spacings $\sim 1$~GeV, i.e. for lattice spacings 
$\sim 0.2$~fm, this symmetry breaking is quite large. In the chiral 
($m_q \rightarrow 0$) limit $m^2_{\pi_2}/m^2_\rho \approx 0.5$ rather than $0$,
a $50$\% effect. Since in the real world, $m^2_{\pi_2}/m^2_\rho \approx 0.03$,
this is a major impediment to working at such lattice spacings. 

For our measurements we have chosen to work with quenched gauge field
configurations at $\beta=5.7$ where the inverse lattice spacing is $\sim 1$~GeV.
To search the parameter space we performed spectrum calculations on 203
independent quenched configurations generated using the standard (Wilson)
gauge action on an $8^3 \times 32$ lattice. Hadron spectra were calculated 
using a single wall source on the $(odd, odd, odd)$ sites of the first time
slice of each configuration, gauge fixed to coulomb gauge. The propagators for
local hadrons were measured using point sinks.

Our measurements used the standard staggered quark action, the MILC action and
our improved action. The $\pi$, $\pi_2$, $\rho$ and nucleon masses are given in
table~\ref{tab:mass8}. For the MILC action we measured the spectrum for
$\omega=0.5$, $\omega=1.0$ and $\omega=\infty$ at quark mass $m_q=0.012$,
$0.02$ and $0.04$. From table~\ref{tab:mass8}, we note that the flavour
symmetry breaking measured as $(m^2_{\pi_2}-m^2_\pi)/m^2_\rho$ appears smallest 
for the $\omega=1.0$ measurements, but that the difference between the 3
$\omega$ values is small. From this we conclude that $\omega=1.0$ is a close
to optimal choice for $\beta=5.7$. For our improved action, even with our
restricted parameterization, there are 6 independent parameters. To limit our
choices, we chose a 1 parameter subclass parameterized by $x$, for which 
$x_n=x^n$ and $y_n=x^{n+1}$. This choice was influenced by tadpole improvement.
The results we present here are for the tree level coefficients ($x=1/2$) with
quark masses $m_q=0.012$ and $0.02$, for $x=1$ with quark masses $0.012$, 
$0.02$ and $0.04$ and for $x=\infty$ with quark masses $0.006$ and $0.012$.
Here symmetry breaking appears to be smallest for $x=1$. We note also that
symmetry breaking for our best improved action is appreciably smaller than for
the best MILC action. Since the $\rho$ masses are very close, this does not
appear to be simply due to differences in the perceived lattice spacing.
\begin{table}[htb]
\begin{tabular}{llllll}
ACTION   & $m_q$ & $m_\pi$    & $m_{\pi_2}$ & $m_\rho$  & $m_N$      \\
\hline
Standard & 0.012 & 0.3156(10) & 0.659(37)   & 0.891(20) & 1.217(22)  \\
Standard & 0.020 & 0.4001(9)  & 0.709(21)   & 0.916(12) & 1.291(15)  \\
Standard & 0.040 & 0.5495(8)  & 0.858(18)   & 1.006(11) & 1.548(31)  \\

MILC5    & 0.012 & 0.3227(32) & 0.462(17)   & 0.793(12) & -------    \\
MILC5    & 0.020 & 0.4083(32) & 0.533(11)   & 0.809(8)  & 1.181(52)  \\
MILC5    & 0.040 & 0.5550(16) & 0.674(8)    & 0.893(8)  & 1.378(16)  \\

MILC1    & 0.012 & 0.3267(33) & 0.453(15)   & 0.787(11) & -------    \\
MILC1    & 0.020 & 0.4100(29) & 0.530(11)   & 0.807(8)  & 1.177(48)  \\
MILC1    & 0.040 & 0.5629(16) & 0.664(6)    & 0.893(8)  & 1.377(15)  \\

MILCu    & 0.012 & 0.3371(32) & 0.471(13)   & 0.781(10) & -------    \\
MILCu    & 0.020 & 0.4246(28) & 0.544(10)   & 0.807(7)  & 1.198(43)  \\
MILCu    & 0.040 & 0.5852(15) & 0.688(5)    & 0.904(7)  & 1.384(20)  \\

naive    & 0.012 & 0.3562(32) & 0.453(9)    & 0.788(9)  & 1.170(15)  \\
naive    & 0.020 & 0.4471(30) & 0.525(6)    & 0.809(7)  & 1.214(11)  \\

imp2     & 0.012 & 0.3868(33) & 0.456(7)    & 0.786(8)  & 1.177(13)  \\
imp2     & 0.020 & 0.4855(22) & 0.538(5)    & 0.815(6)  & 1.264(16)  \\
imp2     & 0.040 & 0.6707(18) & 0.712(3)    & 0.921(5)  & 1.469(40)  \\

impu     & 0.006 & 0.3660(23) & 0.448(8)    & 0.770(10) & 1.163(14)  \\
impu     & 0.012 & 0.5018(22) & 0.554(4)    & 0.813(6)  & 1.328(23)  \\
\end{tabular}
\caption{Hadron masses for various choices of the staggered quark action at
$\beta=5.7$ on an $8^3 \times 32$ lattice. Standard is the standard staggered
action, MILC5, MILC1 and MILCu are the MILC action with $\omega=0.5$, $1$ and
$\infty$ respectively and naive, imp2 and impu are our improved action with
$x=0.5$, $1$ and $\infty$ respectively. 
\label{tab:mass8}}
\end{table}

Since, even at $\beta=5.7$, $8^3$ is a rather small spatial lattice, we have
confirmed and quantified our results using a set of 158 quenched configurations
on a $16^3 \times 32$ lattice, also at $\beta=5.7$. The larger lattice also
permitted us to go to smaller quark masses. For this larger lattice we have
calculated the light hadron spectrum with the standard staggered quark action,
the MILC action with $\omega=1$ and our improved action with $x=1$. For these
measurements we used a single source on time-slice 1 of each configuration.
This source was a constant for all $(odd, odd, odd)$ sites of an 
$8 \times 8 \times 8$ cube and zero elsewhere, making it identical to the
source we used on the smaller lattice, since this seemed to produce flat
effective mass plots. Again we worked in Coulomb gauge and used point sinks.
The masses we obtained from fits to these propagators are presented in
table~\ref{tab:mass16}. 
\begin{table}[htb]
\begin{tabular}{llllll}
ACTION   & $m_q$ & $m_\pi$   & $m_{\pi_2}$ & $m_\rho$  & $m_N$      \\
\hline
Standard & 0.006 & 0.2256(5) & 0.673(23)   & 0.881(10) & 1.346(14)  \\
Standard & 0.012 & 0.3145(4) & 0.723(14)   & 0.909(7)  & 1.392(10)  \\
Standard & 0.020 & 0.4000(4) & 0.813(21)   & 0.941(5)  & 1.442(9)   \\
Standard & 0.040 & 0.5503(4) & 0.909(11)   & 1.026(5)  & 1.551(7)   \\

MILC1    & 0.006 & 0.2233(6) & 0.415(12)   & 0.795(23) & 1.152(34)  \\
MILC1    & 0.012 & 0.3122(6) & 0.458(6)    & 0.819(13) & 1.194(16)  \\
MILC1    & 0.020 & 0.3993(5) & 0.521(4)    & 0.850(8)  & 1.255(12)  \\
MILC1    & 0.040 & 0.5571(5) & 0.656(3)    & 0.909(5)  & 1.394(8)   \\

imp2     & 0.006 & 0.2681(6) & 0.377(6)    & 0.763(15) & 1.111(18)  \\
imp2     & 0.012 & 0.3743(6) & 0.454(4)    & 0.808(8)  & 1.184(11)  \\
imp2     & 0.020 & 0.4785(6) & 0.538(3)    & 0.841(6)  & 1.254(12)  \\
imp2     & 0.040 & 0.6690(7) & 0.713(2)    & 0.938(3)  & 1.410(9)   \\
\end{tabular}
\caption{Hadron masses for various choices of the staggered quark action at
$\beta=5.7$ on a $16^3 \times 32$ lattice. The notation is as for 
table~\protect\ref{tab:mass8}.
\label{tab:mass16}}
\end{table}

We note that the Goldstone $\pi$ masses show very little finite size effect in
going from an $8^3 \times 32$ to a $16^3 \times 32$ lattice. The $\pi_2$ masses
for the MILC and improved actions also show relatively little finite size
effect while that for the standard action shows considerably more. However, we
note that the errors for the standard action are large, and all the errors in
these tables are purely statistical --- no estimate of the systematic error
associated with choice or appropriateness of fits is included --- so that is
not clear how significant this is. Similar comments can be made about any
apparent finite size effects in the $\rho$ and nucleon masses. We assume that
for a $16^3 \times 32$ lattice, where the spatial box size is $> 3$~fm at this
$\beta$, the finite size effects will be relatively small. We refer the reader
to recent, more extensive quenched spectrum calculations at $\beta=5.7$
for serious finite
size studies and masses with which our standard action masses can be compared,
and also for chiral extrapolation studies \cite{milc2,gottlieb}.

Just comparing $(m^2_{\pi_2}-m^2_\pi)/m^2_\rho$ indicates that both the MILC and
our improved actions give a considerable reduction in flavour symmetry 
violation over the standard action. In addition our improved action gives
improvement over that of the MILC action. To make this more quantitative, we
have chirally extrapolated our masses to zero quark mass. We do this firstly
because, since the relationship between the lattice quark mass and the physical
($\overline{MS}$) mass is different for each action, and there is the ambiguity
as to which observable should be used to determine which lattice quark masses
correspond to one another. $m_q=0$ is the same for each lattice action. 
Secondly, the physical $u$ and $d$ quark masses are small enough that the
chiral limit is a good approximation to the real world.

\begin{figure}[htb]
\epsfxsize=5in
\epsffile{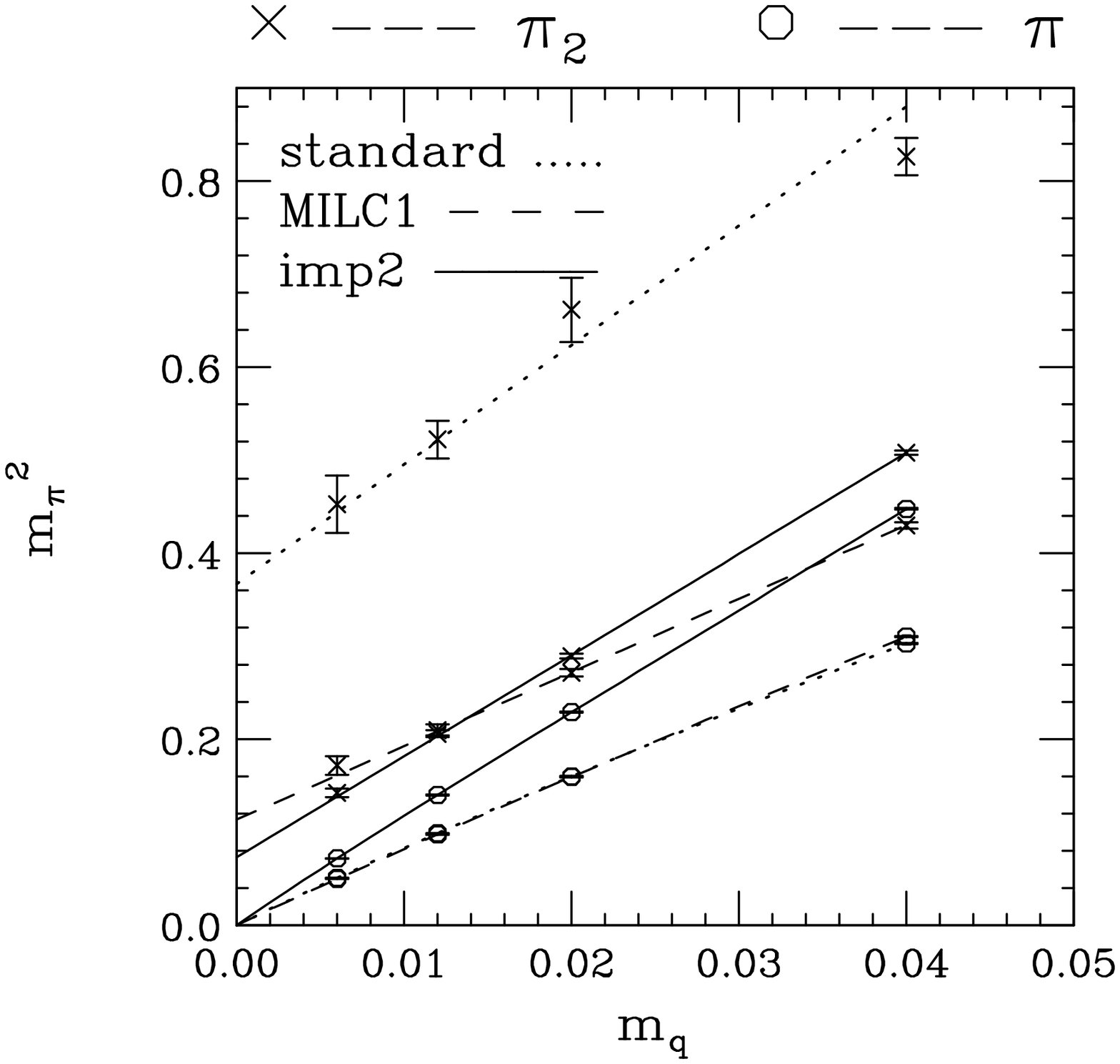}
\caption{Pion masses squared as functions of the quark mass on a 
$16^3 \times 32$ lattice at $\beta=5.7$. The lines are the fits described in
the text.\label{fig:pimass}}
\end{figure}

Since we have 4 quark masses for each action, we can use a 3-parameter fit for
our chiral extrapolation. For the Goldstone pion we have chosen to fit to
\begin{equation}
m_\pi^2 = a m_q + b m_q^{3/2}+c m_q^2 .
\end{equation}
For the second pion $\pi_2$ we have used a 2-parameter fit
\begin{equation}
m_{\pi_2}^2 = a + b m_q .
\end{equation}
Our pion masses and these fits are plotted in figure~\ref{fig:pimass}. Since
our Goldstone pion masses have been forced to zero in the $m_q=0$ limit, 
$m_{\pi_2}^2$ is a good measure of flavour symmetry violation. From our fits
we obtain $m_{\pi_2}^2=0.367(44)$ for the standard action $m_{\pi_2}^2=0.114(8)$
for the MILC action and $m_{\pi_2}^2=0.073(5)$ for our improved action. These
would be valid measures of improvement if we consider that the true lattice
spacing is that obtained from some pure gluonic observable such as the string
tension, or the $\rho$ mass from some unknown ``correct'' fermion action. 

\begin{figure}[htb]
\epsfxsize=5in
\epsffile{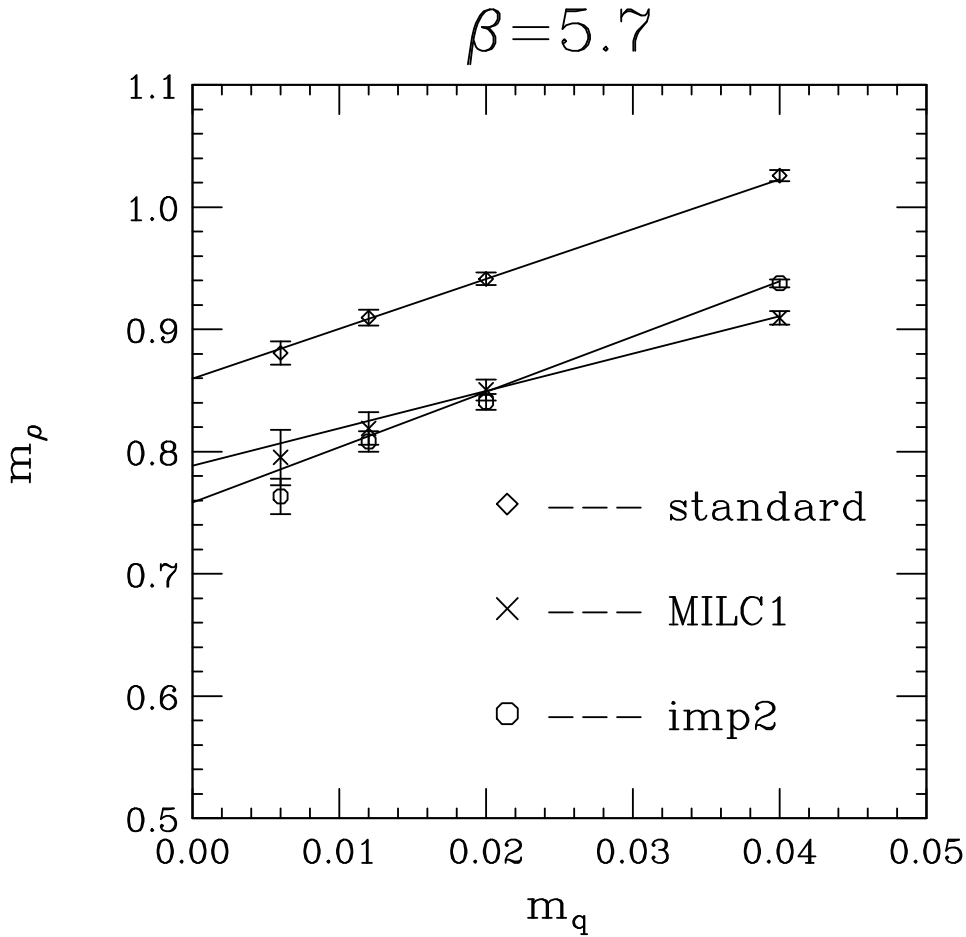}
\caption{$\rho$ masses as functions of $m_q$ for a $16^3 \times 32$ lattice,
with linear fits.\label{fig:rho}} 
\end{figure}

\begin{figure}[htb]
\epsfxsize=5in
\epsffile{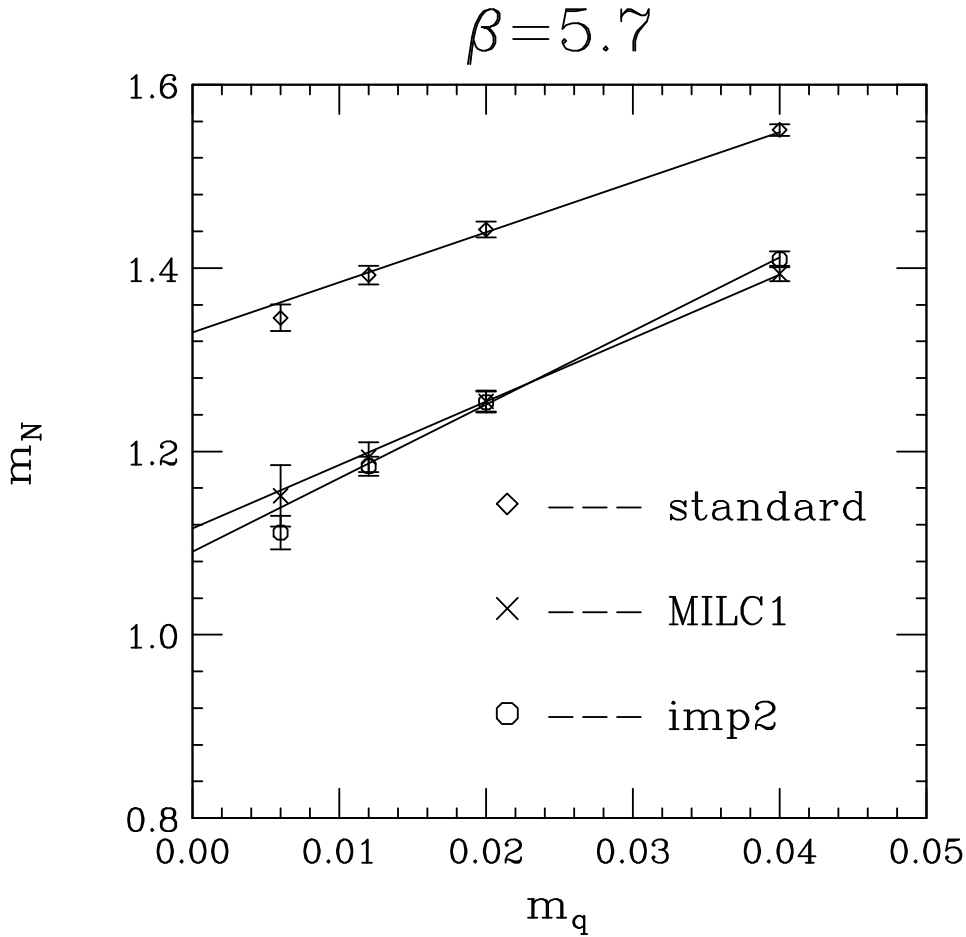}
\caption{Nucleon masses as functions of $m_q$ for a $16^3 \times 32$ lattice,
with linear fits.\label{fig:nucleon}} 
\end{figure}

If, however, we follow the MILC collaboration, and consider that each fermion
action determines its own lattice spacing, we need to extrapolate whatever
hadron mass is to be used to determine this spacing to $m_q=0$ for each action
separately. For the $\rho$ and nucleon, we have used a simple linear 
extrapolation in $m_q$. (Even though such fits were not great, we were unable
to find any 3 parameter fits which did significantly better.) These fits for
our 3 actions are shown in figures~\ref{fig:rho} and \ref{fig:nucleon}. This
gives $m_\rho=0.859(9)$ and $m_N=1.330(14)$ for the standard action, 
$m_\rho=0.788(14)$ and $m_N=1.116(20)$ for the MILC action and 
$m_\rho=0.758(10)$ and $m_N=1.090(17)$ for our improved action in the chiral
limit. Hence if we use the $\rho$ mass to set the scale, we get 
$m^2_{\pi_2}/m^2_\rho=0.497(59)$ for the standard action
$m^2_{\pi_2}/m^2_\rho=0.183(12)$ for the MILC action and
$m^2_{\pi_2}/m^2_\rho=0.127(9)$ for our improved action.

As discussed by the MILC collaboration, one of the effects of improving the
quark action is to reduce the lattice spacing as determined by the $\rho$ mass.
(This is evident in our results quoted above.) Before we conclude how successful
our program has been, we therefore need to know how much flavour symmetry
violation would have been improved merely by making such decreases in lattice
spacing, without changing our quark action. To do this one would need to know
how large the standard action flavour symmetry violations would be at a lattice
spacings where the $\rho$ mass for the standard action would have the values the
MILC $\rho$ mass and our improved $\rho$ mass have respectively at $\beta=5.7$.
In addition, we would need to know the size of the MILC action flavour symmetry
violations at the lattice spacing where the MILC $\rho$ mass has the value
our improved $\rho$ mass has at $\beta=5.7$. We do not have this information.
However, we know that the leading flavour symmetry violations should be
${\cal O}(a^2)$ and the ratios of relevant $a$'s are given by the inverse ratios
of the corresponding $\rho$ masses. This would give an estimate of
$m^2_{\pi_2}/m^2_\rho=0.418(50)$ for the standard action at a lattice spacing
set by the MILC $\rho$ mass, and $m^2_{\pi_2}/m^2_\rho=0.387(46)$ at a lattice
spacing set by our improved $\rho$ mass. Similarly we estimate the flavour
symmetry violation for the MILC action to be $m^2_{\pi_2}/m^2_\rho=0.169(11)$
at a lattice spacing set by our improved $\rho$ mass.

We now check this ${\cal O}(a^2)$ dependence against published results 
\cite{milc} at
higher $\beta$ values, performing the required linear extrapolations as best
we can. At $\beta=5.85$ the chirally extrapolated $\rho$ mass is $0.5676(42)$
from which we predict $m^2_{\pi_2}/m^2_\rho=0.217(26)$ compared with the
value calculated from extrapolated $\pi_2$ and $\rho$ masses namely 
$m^2_{\pi_2}/m^2_\rho=0.267(6)$. At $\beta=5.95$ the chirally extrapolated
$\rho$ mass is $0.4629(40)$ from which we predict 
$m^2_{\pi_2}/m^2_\rho=0.144(17)$ compared with the direct extrapolation
$m^2_{\pi_2}/m^2_\rho=0.170(13)$. Thus we conclude that our predictions from
the assumed ${\cal O}(a^2)$ dependence will be if anything lower than the
actual values.

\section{Discussion and Conclusions}

We have introduced a new class of single-link actions for staggered fermions
which reduce the flavour symmetry violations from ${\cal O}(a^2)$ to 
${\cal O}(a^4)$ at tree level, where $a$ is the lattice spacing, by suppressing
the coupling of high momentum gluons to quarks which is responsible for flavour
mixing. On quenched configurations at $\beta=5.7$ where $a^{-1} \approx 1$~GeV,
the flavour symmetry violations for local pions are reduced by $\approx
65$~--~$75$\%, over those of the standard action and by $\approx 25$~--~$30$\%
over those in the MILC action. For our improved action this means that the
flavour symmetry violations at $\beta=5.7$ are approximately the size of those
for the standard action at $\beta=6.0$, i.e. at approximately half the lattice
spacing. Since all these actions are single link, inverting the Dirac operator
is no more expensive than with the standard action. In fact, it is considerably
less expensive with the improved actions, since they require many less
conjugate gradient iterations to reach the same level of convergence. For
example, at $m_q=0.006$, our improved action required 1000--1050 conjugate
gradient iterations compared with 1700--1750 for the MILC action and 2600--2700
for the standard action. 

Of course, although flavour symmetry violations are one of the most important
barriers to using staggered quarks at lattice spacings of $\gtrsim 0.1$~fm,
they are not the only barriers. Our improvements need to be combined with 
improvements to the free fermion dispersion relations and to the gauge action.
In the case of the MILC action, work of this nature has recently been done by
Orginos and Toussaint \cite{ot}, who have also included dynamical quarks.

The Originos-Toussaint paper does, however, point out that although their
actions improve the $\pi$~--~$\pi_2$, mass splitting, the point-split pions
do not show as much improvement. Although we have not measured the spectrum
of these point-split pions, we expect the improvement to be more uniform
across the pion multiplet with our action than with the MILC action. The
reason is that at tree level, our action uniformly suppresses all flavour
mixings. On the other hand the MILC action, at tree level, can be adjusted to
maximally suppress some flavour mixings, but it will at best only partially
suppress the others. However, only explicit measurement will tell if our
expectations are correct.

Another aspect of flavour symmetry violation for staggered quarks is the extent
to which they fail to obey the Atiyah-Singer index theorem. We have shown that
the MILC action produces only limited improvement in this area \cite{kls}.
It is to be
hoped that our improved action might fare better.

A more serious study is needed to determine the optimal parameters in our
action. Our action is still far from ideal for treating light $u$ and $d$ 
quarks. A tadpole improved perturbative calculation of the coefficients might
be helpful, although perturbative calculations for staggered fermions have
proved disappointing in the past. One might hope that it might be possible to
reduce the flavour symmetry violations to ${\cal O}(a^4)$ as suggested by the
tree level calculations, within the restrictions of a single link fermion
action. The analysis of Luo \cite{luo} should be helpful in reducing the
operators in our action to an independent set.

Although it is obvious that we need to reduce the coupling of high momentum
gluons to staggered quarks, to reduce flavour symmetry violations, it is also
important to reduce such coupling for Wilson quarks. In the case of Wilson
quarks this is to decrease chiral symmetry violations. This has been addressed
most recently by DeGrand \cite{degrand}
who introduced smeared link fields into Wilson fermions
calculations. Such smearing could also be useful in reducing effects of small
instantons on Wilson fermions which have the potential for creating problems
if they are to be used as the basis for domain-wall fermions \cite{ehn}.

\section*{Acknowledgements}
This work was supported by U.S. Department of Energy contract W-31-109-ENG-38.
The computing was performed on the CRAY J-90's at NERSC.


\begin{thebibliography}{999}
\bibitem{niedermayer} F.~Niedermayer, Nuclear Physics B(Proc. Suppl.) {\bf 53},
56 (1997).
\bibitem{symanzik} K.~Symanzik, Nucl. Phys. {\bf B226}, 187 (1983); 
{\it ibid} 205.
\bibitem{lm} G.~P.~Lepage and P.~B.~Mackenzie, Phys. Rev. {\bf D48}, 2250 
(1993).
\bibitem{luscher} S.~Capitani, {\it et al.}, Nucl. Phys. B(Proc. Suppl.) 
{\bf 63}, 153 (1998), and references contained therein.
\bibitem{hasenfratz} P.~Hasenfratz, Nuclear Physics B (Proc. Suppl.) {\bf 63},
53 (1998).
\bibitem{lw}
M.~L\"{u}scher and P.~Weisz, Comm. Math. Phys. {\bf 97}, 19 (1985); Phys. Lett.
{\bf 158B}, 250 (1985).
\bibitem{sw} B.~Sheikholeslami and R.~Wohlert, Nucl. Phys. {\bf B259}, 572
(1985).
\bibitem{alford} M.~Alford {\it et al.}, Phys. Lett. {\bf B361}, 87 (1995);
Nucl. Phys. {\bf B496}, 377 (1997).
\bibitem{degrand} T.~DeGrand, e-print hep-lat/9802012 (1998).
\bibitem{naik} S.~Naik, Nucl. Phys. {\bf B316}, 238 (1989).
\bibitem{luo} Y.~Luo, Phys. Rev. {\bf D57}, 265 (1998).
\bibitem{bbcw} W.~Bietenholz {\it et al.}, Nucl. Phys. {\bf B495}, 285 (1997).
\bibitem{bd} W.~Bietenholz and H.~Dilger, e-print hep-lat/9803018 (1998), 
and references contained therein.
\bibitem{milcs} C.~Bernard, {\it et al.}, e-print hep-lat/9712010 (1997).
\bibitem{bielefeld} A.~Peikert, {\it et al.}, Nucl. Phys. B(Proc. Suppl.)
{\bf 63}, 895 (1998).
\bibitem{milc} T.~Blum, {\it et al.}, Phys. Rev. {\bf D55}, R1133 (1997).
\bibitem{ls} J.-F.~Laga\"{e} and D.~K.~Sinclair, Nucl. Phys. B(Proc. Suppl.)
{\bf 63}, 892 (1998).
\bibitem{milc2} C.~Bernard, {\it et al.}, hep-lat/9805004 (1998);
Nucl. Phys. B(Proc. Suppl.) {\bf 47}, 345 (1996); {\bf 53} 212 (1997).
\bibitem{gottlieb} S.~Gottlieb, Nucl. Phys. B(Proc. Suppl.) {\bf 53}, 155 
(1997), and references contained therein.
\bibitem{ot} K.~Orginos and D.~Toussaint, e-print hep-lat/9805009 (1998).
\bibitem{kls} J.~B.~Kogut, J.-F.~Laga\"{e} and D.~K.~Sinclair, e-print
hep-lat/9801020 (1998).
\bibitem{ehn} R.~G.~Edwards, U.~M.~Heller and R.~Narayanan, e-print
hep-lat/9801015 (1998); R.~G.~Edwards {\it et al.} Nucl. Phys. {\bf B518}, 319 
(1998).

\end{thebibliography}
\end{document}